# A Predictive Analytic Model for High-Performance Tunneling Field-Effect Transistors Approaching Non-Equilibrium Green's Function Simulations

*Ramon B. Salazar [§,\*], Hesameddin Ilatikhameneh[†,\*], Rajib Rahman[†], Gerhard Klimeck[†], and Joerg Appenzeller[§]*

§ Birck Nanotechnology Center, Purdue University, 1205 W. State Street, West Lafayette, IN 47907, IN, USA.

† Network for Computational Nanotechnology, 207 S. Martin Jischke Drive, West Lafayette, IN 47907, IN, USA.

\* These two authors contributed equally to this work.

Keywords: *Tunneling-Field Effect Transistor, Tunneling, Tunnel-FET, Analytic Model, NEGF, ballistic, quantum transport.*

AUTHOR INFORMATION

Corresponding Authors:

E-mail: ramon@purdue.edu, hilatikh@purdue.edu

**Abstract.-**

A new compact modeling approach is presented which describes the full current-voltage (I-V) characteristic of high-performance (aggressively scaled-down) tunneling field-effect-transistors (TFETs) based on homojunction direct-bandgap semiconductors. The model is based on an analytic description of two key features, which capture the main physical phenomena related to TFETs: 1) the potential profile from source to channel, and 2) the elliptic curvature of the complex bands in the bandgap region. It is proposed to use 1D Poisson's equations in the source and the channel to describe the potential profile in homojunction TFETs. This allows to quantify the impact of source/drain doping on device performance, an aspect usually ignored in TFET modeling but highly relevant in ultra-scaled devices. The compact model is validated by comparison with state-of-the-art quantum transport simulations using a 3D full band atomistic approach based on Non-Equilibrium Green's Functions (NEGF). It is shown that the model reproduces with good accuracy the data obtained from the simulations in all regions of operation: the on/off states and the n/p branches of conduction. This approach allows calculation of energy-dependent band-to-band tunneling currents in TFETs, a feature that allows gaining deep insights into the underlying device physics. The simplicity and accuracy of the approach provides a powerful tool to explore in a quantitatively manner how a wide variety of parameters (material-, size- and/or geometry-dependent) impact the TFET performance under any bias conditions. The proposed model presents thus a practical complement to computationally expensive simulations such as the 3D NEGF approach.

**1- Introduction**

Tunneling field-effect-transistors (TFETs) are considered one of the attractive device concepts in the field of nanoelectronics[1]. TFETs have the potential ability to perform with an inverse subthreshold slope ($S$) smaller than conventional thermionic emission-based devices that exhibit S-values above 60mV/decade at room-temperature. This advantage translates in particular into low-voltage operation and may – for the right choice of band gap material – result in low stand-by power dissipation, making TFETs a more energy-efficient device if compared to complementary metal-oxide-semiconductor (CMOS) FETs [1–7].

However, in order to be able to compete with CMOS devices, TFETs must be carefully designed not only to operate with a small $S$, but also to show high on-state currents ($I_{on}$) [1]. The development of analytical models for TFETs is of great importance in this context since those models allow optimizing the transistor design by gaining an in-depth understanding of how to tune the key parameters to enable high performance devices.

In this work, it is shown that in the case of homojunction TFETs whose channel material consists of direct bandgap semiconductors, the current vs. channel-potential characteristics of the device can be described by combining analytic models which capture: a) the total band bending distance from source to channel (Λ) which takes into account the impact of doping in the source/drain regions, and b) the elliptic curvature of the complex band structure, which describes the tunneling of electrons and holes through the band gap. It is shown in particular that aggressively scaled down TFETs for high-performance applications are impacted more by the depletion width in the source/drain regions than by the geometric screening length, which is in turn captured by the analytic expressions. The proposed model provides results, which are in good agreement with those obtained through rigorous numerical calculations using the 3D self-consistent Poisson-Non-Equilibrium Green's Function (NEGF) methodology in our simulation engine NEMO5.

**2.- Theory**.

**2.1 The Effective Screening Length**

Theoretical studies are frequently concerned with homojunction TFETs with a heavily doped source and a lightly doped (or intrinsic) channel [8–13] and assume there is a total *band bending distance* from source to channel denoted as "Λ" (see Fig. 1). Moreover, it is often assumed that the source is doped to such a high extent that there is almost no depletion (or band bending) in this region and thus $\Lambda \approx \lambda_{ch}$ as shown in Fig. 1a ($\lambda_{ch}$ is the *geometric screening length* of the gated semiconductor[14–16]). Previous reports have discussed how neglecting the band bending inside the source/drain regions leads to an overestimation of the electric field at the source/channel (and channel/drain) junction[17]. In this work it will be shown that for high-performance devices (ultra scaled-down) where $\lambda_{ch}$ is very small, neglecting the band bending in the source/drain regions indeed leads to significantly overestimated on-currents of band-to-band (BTB) TFETs.

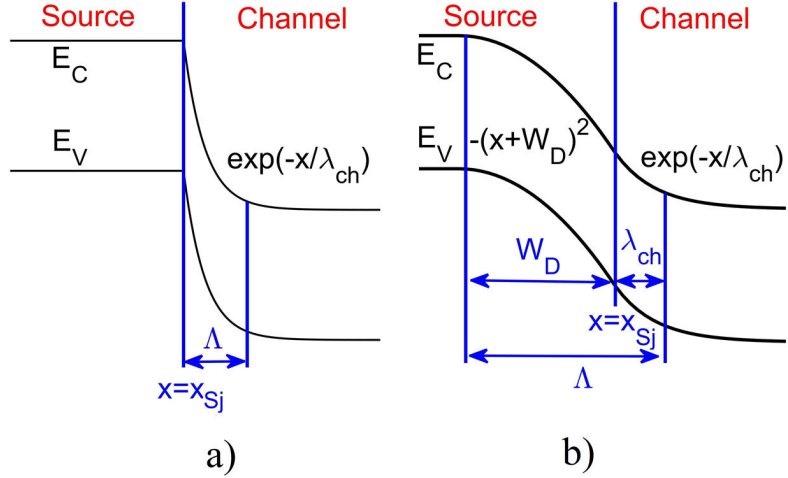

**Figure 1.** Schematics of the potential profile along the channel of an ultra-scaled homojunction TFET assuming **a)** infinite doping level in the source and **b)** finite doping in the source. Notice that $W_D \gg \lambda_{ch}$. The point $x = x_{sj}$ defines the source/channel interface.

A more appropriate description of the electrostatics in homojunctions leads to a potential profile as shown in Fig. 1b. In this case, there is a band bending distance associated with both the source (i.e., the depletion width "$W_D$"), and the channel ($\lambda_{ch}$). It will be shown later that in aggressively-scaled down TFETs usually $W_D \gg \lambda_{ch}$ as shown in Fig. 1b.

**2.2 Analytic Description of the Potential Profile**

Finding an analytic solution to the full 3D-Poisson's equation from source to channel to describe the potential profile and the electric field can be a non-trivial task. Instead, a different approach is proposed to solve this problem in a simpler way. For the first time, it is proposed to describe the potential profile in homojunction TFETs using 1D Poisson's equations in the source and the channel. In the depleted region of the source far from the source-channel interface the Poisson's equation can be written as:

$$\frac{d^2U}{dx^2} = -\frac{qN_D}{\epsilon}.$$

(1)

where $q$ is the elementary charge of the electron, and $U$, $N_D$, and $\epsilon$ are: the potential (in the source), the corresponding doping concentration in the source (or drain), and the in-plane permittivity of the semiconductor in that region respectively. In contrast, far from the source/channel interface the Poisson's equation for the channel potential region can be written as[14]:

$$\frac{d^2U}{dx^2} = \frac{U}{\lambda_{ch}^2}.$$

(2)

### 2.2.1 Piecewise Analytic potential

The general forms of the solutions to (1) and (2) are given by $U(x) = A(x - x_{Sj} + W_D)^2$ in the source, and $U(x) = (\phi_j - \phi_{ch})exp\left(-\frac{x-x_{Sj}}{\lambda_{ch}}\right) + \phi_{ch}$ in the channel, where A is a constant and $W_D$ is the depletion width in the source region, $\phi_j$ is the potential at the source/channel junction, and $\phi_{ch}$ is the potential in the channel as shown in Fig. 2b and 2c. Furthermore, it is assumed that these solutions are still valid in the region close to the source/channel interface. In section A.1 of the Appendix it is shown that (1) and (2) lead to a continuous piecewise analytic potential profile from source to channel denoted as $U_P(x)$:

$$U_P(x) \approx \begin{cases} \phi_s & x \leq x_{Sj} - W_D \\ \phi_s - \frac{\frac{W_D}{2}}{\lambda_{ch} + \frac{W_D}{2}}(\phi_s - \phi_{ch})\frac{(x - x_{sj} + W_D)^2}{W_D^2} & x_{Sj} - W_D < x < x_{Sj} \\ \phi_{ch} + \frac{\lambda_{ch}}{\lambda_{ch} + \frac{W_D}{2}}(\phi_s - \phi_{ch})exp\left(-\frac{x - x_{sj}}{\lambda_{ch}}\right) & x_{Sj} \geq x \end{cases}$$

(3)

## 2.2.2 Ad-hoc Analytic Potential

It is also possible to find a simpler description of the potential profile from source to channel that captures accurately the impact of $\lambda_{ch}$ and $W_D$ on the tunneling distance. The ad-hoc analytic potential shown in (4) is proposed as a simple and adequate solution without resorting to piecewise functions. Although (4) is an arbitrary choice (to some extent) all the parameters contained in the equation are well defined and thus there are no fitting parameters that have to be unjustifiably tuned. Therefore, (4) is considered as predictive and it is defined as:

$$U_A(x) = \frac{\phi_S - \phi_{ch}}{1 + exp\left(\frac{x - x_{Sj}}{\lambda_{eff}}\right)} + \phi_{ch}, \tag{4}$$

where $U_A(x)$ is a continuous solution of the potential from source to channel, and $\lambda_{eff}$ is the *potential decay rate* at the source/channel junction (other parameters has been described above). By setting the requirement that close to threshold $U_A(x)$ must describe the same band bending distance ($\Lambda$) as the piecewise approach, the decay rate $\lambda_{eff}$ can be shown to be (see details in appendix A.2):

$$\lambda_{eff} = \frac{W_D + \lambda_{ch}}{6}, \tag{5a}$$

Where

$$\lambda_{ch} = \sqrt{\frac{\varepsilon_{body}}{\varepsilon_{ox}} \frac{t_{body}^2}{8} \ln\left(1 + \frac{2 t_{ox}}{t_{body}}\right) + \frac{t_{body}^2}{16}}, \tag{5b}$$

$$W_D \approx \sqrt{\frac{2\epsilon}{q N_D} \alpha (\phi_S - \phi_{ch})}. \tag{5c}$$

In (5b) $\varepsilon_{body}$ and $\varepsilon_{ox}$ are the permittivities of the semiconductor and oxide respectively, and $t_{body}$ and $t_{ox}$ are the diameter of the nanowire and oxide thickness of the cylindrical structure respectively. In (5c) $W_D$ is the approximated depletion width at the source (see details in appendix

A.2). Notice that the description of $\lambda_{ch}$ as shown in equation (5b) for cylindrical geometries is accurate for ultra-scaled TFETs where the radius of the nanowire is comparable to the oxide thickness[18]. The conduction and valence band potential profiles from source to drain for a complete TFET device can be thus obtained based on equation (4) as:

$$\Phi_C(x) = \frac{\phi_S - \phi_{ch}}{1 + exp\left(\frac{x - x_{Sj}}{\lambda_{eff}}\right)} + \frac{\phi_{ch} - \phi_D}{1 + exp\left(\frac{x - (x_{Sj} + L_{eff})}{\lambda_{eff}}\right)} + \phi_D, \qquad (6a)$$

$$\Phi_V(x) = \Phi_C(x) - E_g. \qquad (6b)$$

In equation (6) $\Phi_C(x)$ and $\Phi_V(x)$ are the potential profiles of the conduction and valence band respectively along the direction of transport, $L_{eff}$ is the "effective channel length" of the channel defined as $L_{eff} \approx L_{ch} + \delta x_{Sj} + \delta x_{Dj}$, where $L_{ch}$ is the actual channel length of the device, and $\delta x_{Sj}$ ($\delta x_{Dj}$) is the symmetry shift of the potential at the source/channel (drain/channel) junction (see section A.4 of the appendix for details), $E_g$ is the bandgap of the semiconductor, and $\phi_D$ is the potential in the drain (the value at the edge of the conduction band). Notice that $\lambda_{eff}$ as described in (5) does not have to be necessarily the same in the source/channel junction as it is in the channel/drain junction. All other parameters have been previously introduced. Note that the values of $\phi_S$ and $\phi_D$ relative to the Fermi Levels in the source/drain (respectively) depend on the doping concentration in those regions and can be found from well-known Fermi-Dirac integrals[19]. Figure 2 summarizes the details of the proposed piecewise and ad-hoc potential profiles.

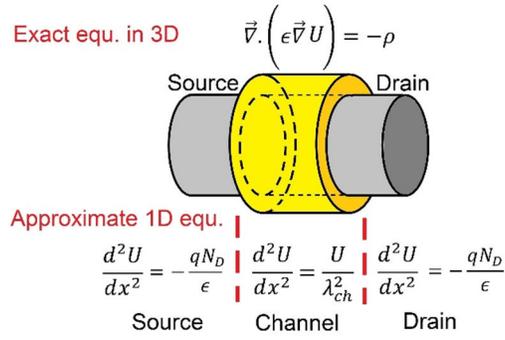
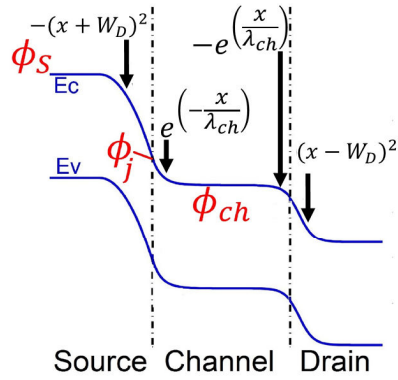
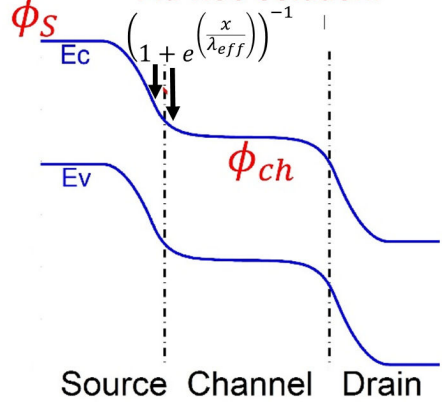

**Figure 2**. **a)** Illustration of a nanowire TFET showing the 1-D Poisson equations proposed in each region of the device as shown in (1) and (2). Next to the illustration are found band diagrams of the device in "**a**" showing the general form of: **b)** solutions to 1-D Poisson's equation in each region (piecewise potential profile). **c)** the ad-hoc potential profile.

Moreover, it can be proven that according to (4) the maximum electric field at the source/channel junction ($x = x_{sj}$) is given by (see section A.3 of the appendix):

$$E_{Max} = \frac{\phi_S - \phi_{ch}}{4\lambda_{eff}} = \frac{\phi_S - \phi_{ch}}{\frac{4}{6}(W_D + \lambda_{ch})}. \qquad (7)$$

It is important to note the significant difference between the common assumption of having a maximum electric field $E'_{MAX} = \frac{\phi_S - \phi_{ch}}{\lambda_{ch}}$ and the result shown in (7). In particular, equation (7) supports the argument presented earlier: "When $W_D \geq \lambda_{ch}$ (which is the case for drastically scaled devices where $\lambda_{ch}$ is small) the assumption of $W_D$ being negligible is incorrect". More specifically, assuming $W_D \approx 0$ leads to an overestimation of the electric field ($E'_{MAX} \geq \frac{4}{3} E_{Max}$), which ultimately translates into a significant error in the calculation of the BTBT current due to its exponential dependence on the strength of the electric field. Although the effect of doping on homojunction TFET performance has been pointed in previous work[20], to the best of our knowledge this is the first time that the impact of source/drain doping concentrations are properly integrated into a novel analytical approach and validated with state-of-the-art quantum transport simulations. In previous reports dealing with analytic modeling of homojunction TFETs the impact of doping has not been treated properly since it has been either neglected[13,21,22], or avoided by introducing directly a numeric value of the electric field without relating it to a realistic doping concentration[9,12,23–25]. On the other hand, the average electric field given by $E_{avg} = \frac{\phi_S - \phi_{ch}}{6\lambda_{eff}}$ (section A.2 of the appendix) allows to relate the potential decay rate ($\lambda_{eff}$) to the total band bending distance ($\Lambda$) through the observation that $6\lambda_{eff} = \Lambda$ and consequently $\Lambda = W_D + \lambda_{ch}$ (see Fig. 1b).

Notice that although (3) could describe the potential in heterostructure TFETs, it has been shown that electron-phonon scattering impacts the device performance significantly[26–28]. Therefore, the WKB approximation may not provide a realistic description of transport phenomena.

**3- Numerical Methods**

To test the quality of the above analytical potential by comparison with a well-established modeling approach, the self-consistent Poisson-NEGF (Non-Equilibrium Green's Function[29,30]) methodology has been used within the tight binding (TB) description to simulate various TFETs. Simulations have been performed by our nanodevice simulation tool NEMO[31,32]. The TB model uses an $sp^3d^5s^*$ basis with nearest-neighbor interaction to describe InAs and Si Hamiltonians[33]. This TB model captures the band structure of InAs and Si and the effect of confinement accurately[8].

The choice of InAs as a channel material in this work has been made purely on the basis of its convenience in terms of: 1) availability, 2) exhibiting a direct band gap with a single dominant tunneling path, which allows the use of the Wentzel–Kramer–Brillouin (WKB) approximation[8], and 3) relatively low bandgap/effective mass, which leads to high BTBT currents. However, the applicability of the proposed model is not restricted to this material system – in fact a comparison with silicon devices is provided as an example of another material system where is found that the potential profile calculated through the proposed analytical approach is actually in good agreement with the self-consistent simulations. This is shown in section A.6 of the appendix where the results from self-consistent simulations are compared with those obtained from equation (6) in the case of a Si nanowire TFET. This provides support to the claim that equation (6) is applicable to different semiconductors in a general manner.

It is important to point out that because of the choice of low density-of-sates (DOS) InAs channel material the relation between channel-potential and gate-voltage is almost one-to-one. If a different semiconductor with high DOS is chosen or if back-injection from the drain becomes dominant[26,34] the model still allows calculating current vs. channel-potential but a capacitor network (series of oxide, quantum[35–37], and electrostatic/offset[38] capacitances) is required to translate the channel-potential into a gate-voltage.

## 4.- Modeling TFETs

In its most accurate form, the tunneling transmission based on the WKB approximation is expressed as follows:

$$T_{WKB}(E) = exp\left(-2 \int_{x_{ini}}^{x_{fin}} k_{IM}(x,E)dx\right), \tag{8}$$

Where $k_{IM}(x,E)$ is the imaginary part of the complex crystal momentum wavevector ($k=k_{RE}+ik_{IM}$) as a function of energy and position, and "$x_{ini}$" and "$x_{fin}$" are the initial and final position of carriers in real space, respectively, when they tunnel elastically through the bandgap. It is noteworthy that the WKB approximation has been found to be an accurate description of transport phenomena in homojunction TFETs when compared to results from NEGF simulations including both ballistic and electron-phonon scattering[8].

To obtain an accurate analytic expression for $k_{IM}(x,E)$, it is necessary to find an analytical equation for: 1) $k_{IM}(E)$ - which has been shown to be an elliptic curve that can be described using an analytic expression that agrees with rigorous tight-binding simulations within 1.4% error[39] (see section A.5 of the appendix) and 2) the potential profile of the conduction/valence bands from source to drain which is described by equations (3) or (6).

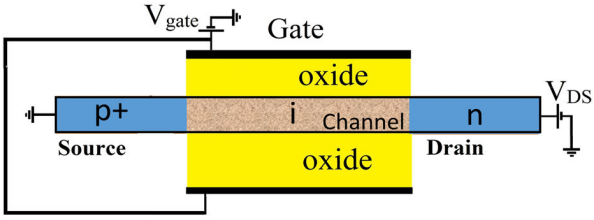
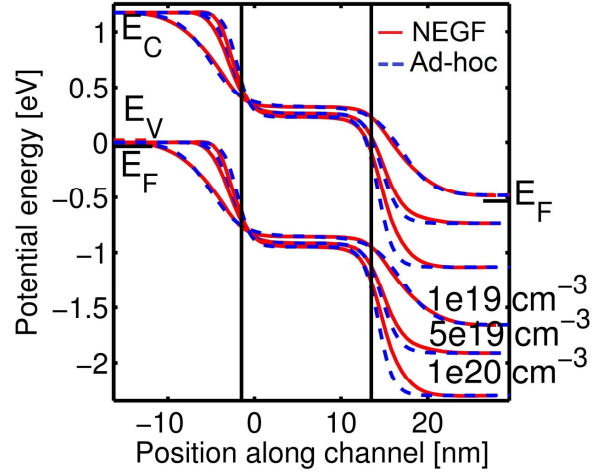

a)

b)

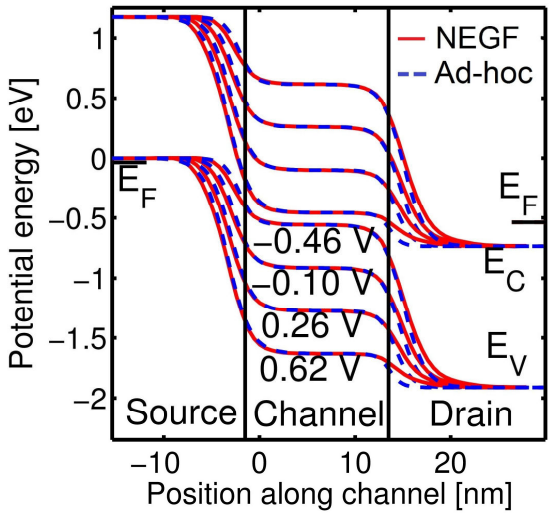
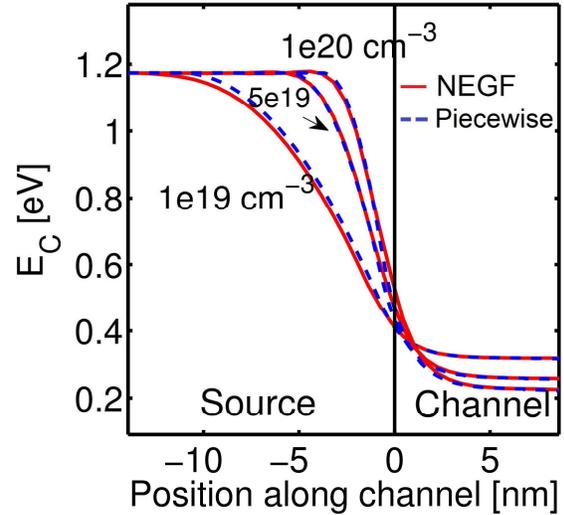

c)

d)

**Figure 3. a)** Cross section of the simulated device: GAA InAs NW TFET. The ad-hoc potential obtained using (6) is compared with state-of-the-art NEGF simulations (red) for: **b)** different source doping concentrations "$N_D$" (source and drain doping levels are set equal) at a constant gate-voltage ($V_G$ =-0.1V), and **c)** different channel potentials at a constant doping ($N_D$ =5x10$^{19}$cm$^{-3}$). The vertical black lines show the interfaces among the source/channel/drain regions. **d)** Piecewise potential profile described by (3) compared to NEGF simulations results for different $N_D$. The applied drain-voltage is $V_D$=0.5V in all cases.

In Figs. 3b and 3c the ad-hoc potential approach using equation (6) is compared with NEGF simulations. In Fig. 3b the comparison is made assuming three different doping concentrations in the source/drain regions (source and drain are equally doped) for the same gate-voltage ($V_G$). In Fig. 3c the comparison is made assuming four different values of $V_G$ while keeping the doping concentration constant. In Fig. 3d the piecewise potential approach using equation (3) is compared with NEGF simulations for three different doping concentrations in the source/drain regions. Notice in Figs. 3b and 3d that for all (high) doping concentrations the distance associated with the band bending inside the source ($W_D$) is greater than the distance associated with the band bending inside the channel ($\lambda_{ch}$). Based on these observations it is clear that accurate calculation of tunneling transmission probabilities in ultra-scaled TFETs (where the band bending inside the gated channel is very small) requires taking into account the band bending (depletion width) inside the highly doped source/drain as well.

Section A.6 of the appendix shows the comparison between the potential profiles obtained through NEGF simulations and those obtained from equation (6) for a GAA Si NW TFET. Figures 3 and A1 suggest that both analytic approaches (ad-hoc and piecewise) have general applicability and they reproduce with good accuracy the potential landscape of different semiconductors for varying $V_G$ and doping concentrations when compared with the results from NEGF simulations. It is important to point out that appropriate comparison of BTBT currents between the proposed analytic approach and NEGF simulations is not possible in Si nanowires since equation (8) (WKB approximation) provides accurate results only when is applied to semiconductors with a single dominant tunneling path through the band gap. However, according to previous work this condition is not satisfied in the case of Si nanowires[8].

The BTBT currents in TFETs can be described using the following analytic model. Using equation (3) or (6) one can write:

$$k_{IM}(x, E_\alpha) = k_{IM}(E = E_\alpha - |q\Phi_V(x)|) , \qquad (9)$$

Where $k_{IM}(E)$ is an analytical expression according to Ref. 39 ($k_{IM}(E)$ is shown in appendix A.5 for the sake of clarity and practicality) and $E_\alpha$ is the value of energy at which carriers tunnel elastically between the conduction and valence bands. Then the BTBT "spectral current" (current per unit energy) is essentially given by combining (8) and (9):

$$I_{spect}(E_\alpha) = \frac{2q}{h} exp\left(-2 \int_{x_{ini}}^{x_{fin}} k_{IM}(x, E_\alpha) dx\right), \qquad (10)$$

where $x_{ini}$ and $x_{fin}$ can be found self-consistently from equations (3) or (6) by satisfying the condition $|q\Phi_V(x_{ini})| = |q\Phi_C(x_{fin})| = E_\alpha$, and $h$ is the Planck's constant. Finally, the total BTBT current is obtained by integrating the spectral current and taking into account the thermal distributions of the Fermi functions:

$$I_{BTBT} = S \cdot \int_{-\infty}^{\infty} I_{spect}(E_\alpha) \left[f_{E,S}(E_\alpha) - f_{E,D}(E_\alpha)\right] dE_\alpha . \qquad (11)$$

Where $S$ is the spin degeneracy, and $f_{E,S}$ and $f_{E,D}$ are the Fermi functions at the source and drain respectively.

The model described by equations (10) and (11) enables to calculate the spectral and total tunneling currents respectively both in the on- and off-state of the device for all bias conditions set by $\phi_{ch}$ and $\phi_D$. The parameters required for the calculation are $m_h$, $m_e$, $E_g$, $L_{ch}$, and $\lambda_{eff}$. The three first parameters are material dependent quantities, which are the *confined* hole and electron effective masses, and the *confined* bandgap of the semiconductor, respectively. The three latter are determined by the scaling approach used in the design (dimensions and geometry of both the channel and oxide, and the doping concentrations in the source and drain). Therefore, this analytic model allows to quantitatively explore how a vast variety of parameters (either material-dependent, size and/or geometric) can impact the TFET performance under any bias conditions. Notice that this model describes many relevant tunneling phenomena in TFETs without having to resort to arbitrary fitting parameters or complex numerical methods such as NEGF with high computational burden.

Notice that since (3), (6) and (A.20) have been shown to be good analytic models for different direct bandgap semiconductors it follows that (10) and (11) should be applicable in general.

## 5.- Results And Comparisons

To validate the analytic model the transfer characteristics (I-$V_G$) calculated through NEGF (solid lines), are compared with those obtained through the piecewise approach by combining (3) and (11) (cross symbols), and those from the ad-hoc approach by combining (6) and (11) (circle symbols) and shown in Fig. 4. The investigated device is the GAA InAs NW TFET described earlier which was simulated to obtain the results presented in Fig. 3. Figure 4a shows the I-$V_G$ for three different doping concentrations (i.e. different $W_D$) while keeping $\lambda_{ch}$ constant ($\lambda_{ch} \approx 1.36 nm$). Figure 4b is a zoom-in of Fig. 4a revealing more clearly the details of the curves in the on-state of the device. Figure 4c shows the comparison between I-$V_G$ curves as the relative oxide dielectric permittivity ($\kappa_{ox}$) is changed from 2.25 ($\lambda_{ch} \approx 2.3$nm) to 36 ($\lambda_{ch} \approx 1$nm) while the source/drain doping concentrations are kept constant.

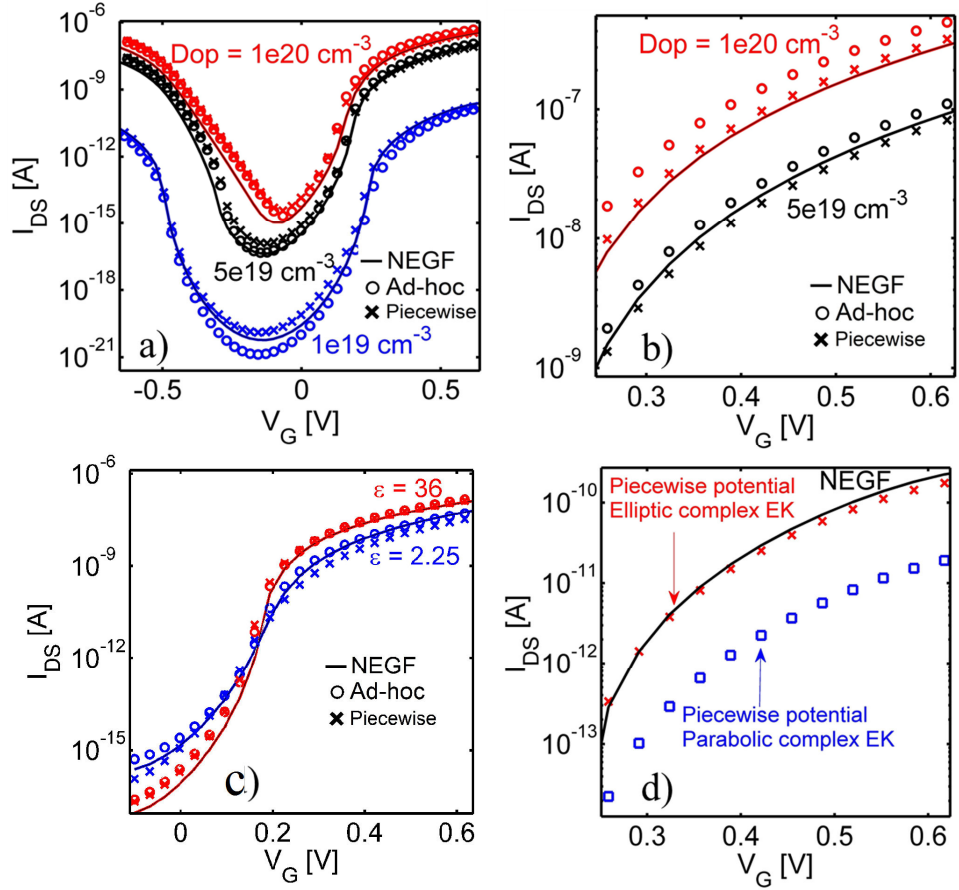

**Figure 4.** Comparison of transfer characteristics for: **a)** three different doping concentrations while keeping $\lambda_{ch}$ constant ($\lambda_{ch} \approx 1.36 nm$), a zoomed in plot of these curves is shown in **b)**, **c)** two different oxide dielectric permittivities (i.e., two different values of $\lambda_{ch}$) while the source/drain doping concentrations are kept constant ($N_D$=5x10$^{19}$cm$^{-3}$), and **d)** NEGF and piecewise analytic approach using elliptic (red) and parabolic (blue) complex bands. The applied drain voltage is $V_D$=1V. The simulated device is a GAA InAs NW TFET (same structure as shown in Fig. 2a).

It is clear that both the piecewise and ad-hoc approaches proposed in this work reproduce with good accuracy the results obtained through NEGF for different values of $W_D$ and $\lambda_{ch}$. In addition, notice that the analytic models are in good agreement with the NEGF simulations both in the on- and off-state of the device and for both the p- as well as the n-branch of conduction. In Figure 4d, the I-$V_G$ obtained from the piecewise approach using elliptic (red) and parabolic (blue) complex bands are compared with NEGF simulations. This illustrates that an analytic approach using a parabolic approximation may underestimate the current significantly, which is consistent with the findings in Ref. 39. This shows that an appropriate representation of 1) the variation of the potential profile along the channel including the impact of doping in the source/drain regions, and 2) the elliptic curvature of the complex bands through the bandgap, provides the means to describe the main physical phenomena related to BTBT in TFETs.

In previous publications an expression based on the Fowler-Nordheim (FN) approach given by $T_{WKB}^{FN} \approx exp\left(-\frac{4\lambda_{ch}\sqrt{2m_{red}^*}E_g^{3/2}}{3\hbar(E_g+\Delta E_{ch})}\right)$ has been frequently used as a means to estimate the BTBT current in TFETs[1,2,13,21,35]. This implies that equation (8) has been simplified by assuming: 1) a constant electric field ($E_0$) in the tunneling window given by $E_0 \approx (E_g + \Delta E_{ch})/q\lambda_{ch}$ where it is assumed that $\Lambda \approx \lambda_{ch}$ ($\Delta E_{ch}$ is the energetic tunneling window), and 2) that carriers tunnel between valence and conduction band with a reduced effective mass given by $m_{red}^* = \frac{m_e m_h}{m_h+m_e}$. Although these assumptions might not be significantly inaccurate in the case of Schottky-Barrier (SB) FETs (metal-semiconductor tunneling junctions)[2,35], in the case of tunneling homojunctions[1,13,21] such assumptions may lead to a significant overestimation of the electric field at the junction, which consequently produces vast errors in the calculation of the BTBT currents. Table 1 shows the comparison of the BTBT on-currents (for the same overdrive voltage) estimated using: 1) the FN approach, 2) the ad-hoc approach, 3) the piecewise approach, and 4) NEGF simulations for the same doping concentrations in the source as shown in Figure 4. Notice that the FN approach severely overestimates the on-current by as much as 4 orders of magnitude for a high doping of $N_D$=1x10$^{19}$cm$^{-3}$. As the doping is further increased up to the level of $N_D$=1x10$^{20}$cm$^{-3}$ the error from using the FN approach becomes less dramatic as expected. It is noteworthy that the results of the analytic model follow the same trends as those from NEGF (e.g. several orders of magnitude change in ON-current due to change in the source/drain doping).

Table 1. Comparison of on-currents calculated through different methods for the same device simulated in Figs. 3 and 4. The gate overdrive voltage and doping concentrations are specified in each case.

|  | ON-CURRENT (A) ($N_D$=1x10$^{19}$cm$^{-3}$) ($V_G$- $V_{Th}$ = 0.4V) | ON-CURRENT (A) ($N_D$=1x10$^{20}$cm$^{-3}$) ($V_G$- $V_{Th}$ = 0.4V) |
|---|---|---|
| **Fowler-Nordheim** | 6.4x10$^{-6}$ | 6.4x10$^{-6}$ |
| **Ad-hoc approach** | 1.3x10$^{-10}$ | 4.6x10$^{-7}$ |
| **Piecewise approach** | 1.75 x10$^{-10}$ | 3.5 x10$^{-7}$ |
| **NEGF** | 2.2x10$^{-10}$ | 3.1x10$^{-7}$ |

Finally, it is instructive to realize that equation (10) provide the means to perform an energy-dependent analysis of BTBT currents in TFETs. Therefore, the proposed model can be used as a powerful resource to gain both qualitative and quantitative insight into the details of how carriers tunnel as a function of energy, depending on the specific bias conditions. In section A.7 of the appendix it is shown how equation (10) allows to inspect in great detail the dynamics of tunneling carriers. This enables e.g. to understand that at low drain bias the off-current "$I_{off}$" of the device: 1) is given by direct tunneling from source to drain through the gated region, and 2) it mimics the function $k_{IM}(E)$ within the bandgap of the semiconductor. These two conditions may lead to a vast deterioration of the inverse subthreshold slope (S), which is one of the most relevant aspects of design in TFETs. The insight that at low drain bias $I_{off}$ resembles the curvature of $k_{IM}(E)$ is particularly important in a voltage-scaling trend where the drain bias ($V_D$) becomes small. In this scenario improving the off-state performance (reducing both S and $I_{off}$) requires the use of large effective mass semiconductors (see appendix A.7). However, this would decrease the on-state current and thus further work is required to find an appropriate solution that improves both the off- and on-state performance in ultra-scaled TFETs operating at low bias.

# 7.- Conclusions

A new analytic approach has been developed for the calculation of BTBT currents in high-performance (aggressively scaled-down) ballistic TFETs which use homojunction direct-bandgap semiconductors with a single dominant tunneling path as channel material. The data calculated through the use of both the piecewise and an ad-hoc approaches are found to be in good agreement with the results obtained from state-of-the-art NEGF simulations in all regions of operation of the device, i.e., the on- and off-state, and the p- as well as the n-branch of conduction. The model is based on appropriate description of: 1) the potential profile along the channel including the impact of doping, and 2) the elliptic curvature of the complex bands through the bandgap, which enable to describe the main physical phenomena related to BTBT in homojunction TFETs. It has been shown that as TFETs are ultra scaled down to improve their performance the depletion width in the source/drain regions has a stronger impact on the TFET performance compared with the effect of the geometric screening length, which is an aspect that has not been taken into account before in TFET modeling. The simplicity and accuracy of the analytic approach makes it attractive to explore how a wide variety of parameters (material-dependent, size and/or geometric) can impact the TFET performance under any bias conditions without having to resort to sophisticated numerical methods such as NEGF, which have high computational complexity. Moreover, the developed compact model allows to perform an energy-dependent analysis of BTBT currents in TFETs, a powerful resource to study the underlying device physics in various TFETs. This type of analysis revealed that at low drain bias the off-current mimics the curvature of $k_{IM}(E)$, which is a relevant insight in a voltage-scaling trend where the drain-voltage becomes small.

**Acknowledgements**

This work was supported in part by the Center for Low Energy Systems Technology (LEAST), one of six centers of STARnet, a Semiconductor Research Corporation program sponsored by MARCO and DARPA. The use of nanoHUB.org computational resources operated by the Network for Computational Nanotechnology funded by the US National Science Foundation under grant EEC-1227110, EEC-0228390, EEC-0634750, OCI-0438246, and OCI-0721680 is gratefully acknowledged.

# APPENDIX A

# Additional Insights And Mathematical Justifications For The Analytic Model

## A.1: Derivation of the Piecewise Potential Profile

In a homojunction structure the continuity of the electric field is already implied because the (in-plane) permittivities in the source and channel regions are equal. Therefore, by making $\frac{\partial U_1}{\partial x}\big|_{x=x_{sj}} = \frac{\partial U_2}{\partial x}\big|_{x=x_{sj}}$ where $U_1$ and $U_2$ are equations (1) and (2) (respectively), and by noting that $U_2(x_{sj}) = A(W_D)^2 = (\phi_j - \phi_s)$ ($\phi_s$ is the potential at the source as shown in Fig. 2b-c) the following expression can be obtained:

$$\frac{(\phi_j - \phi_s)}{\frac{W_D}{2}} = -\frac{(\phi_j - \phi_{ch})}{\lambda_{ch}}. \tag{A.1}$$

Solving for $\phi_j$ in (A.1) allows to find expressions for A and $(\phi_j - \phi_{ch})$:

$$A = \frac{\frac{W_D}{2}}{\lambda_{ch} + \frac{W_D}{2}} \frac{(\phi_s - \phi_{ch})}{W_D^2}, \tag{A.2}$$

$$(\phi_j - \phi_{ch}) = \frac{\lambda_{ch}}{\lambda_{ch} + \frac{W_D}{2}} (\phi_s - \phi_{ch}). \tag{A.3}$$

In this way, the solutions to (1) and (2) are rewritten using (A.2) and (A.3) which leads to the full continuous piecewise analytic potential profile from source to channel denoted as $U_P(x)$:

$$U_P(x) \approx \begin{cases} \phi_s & x \leq x_{Sj} - W_D \\ \phi_s - \dfrac{\frac{W_D}{2}}{\lambda_{ch} + \frac{W_D}{2}}(\phi_s - \phi_{ch})\dfrac{(x - x_{Sj} + W_D)^2}{W_D^2} & x_{Sj} - W_D < x < x_{Sj} \\ \phi_{ch} + \dfrac{\lambda_{ch}}{\lambda_{ch} + \frac{W_D}{2}}(\phi_s - \phi_{ch})\exp\left(-\dfrac{x - x_{Sj}}{\lambda_{ch}}\right) & x_{Sj} \geq x \end{cases}$$

(A.4)

## A.2: Decay Rate of The Ad-hoc Potential Profile

In this section we consider the ad-hoc solution for the potential profile in the source-to-channel region:

$$U(x) = \frac{\phi_s - \phi_{ch}}{1 + \exp\left(\frac{x - x_{Sj}}{\lambda_{eff}}\right)} + \phi_{ch}. \tag{A.5}$$

where $x_{Sj}$ is the position of the source/channel interface, $\phi_S$ and $\phi_{ch}$ are the source and channel potentials far from $x_{Sj}$, and $\lambda_{eff}$ is the *potential decay rate* at the source/channel junction. Notice that $U(x)$ as shown in (A.5) has first and second derivatives which are continuous at the junction ($x = x_{Sj}$). The parameter $\lambda_{eff}$ has to be chosen such that it keeps consistency with the underlying physics of the structure. Next, it is discussed how to choose the appropriate value for $\lambda_{eff}$:

It has already been established in previous work [17] that the solution to Poisson's equation inside the gated region can be written as:

$$U(x) = (\phi_j - \phi_{ch})\exp\left(-\frac{x - x_{Sj}}{\lambda_{ch}}\right) + \phi_{ch}, \tag{A.6}$$

where $\phi_j$ is the potential at the source/channel junction. Consider the change in potential "$U'$" defined as $U' = U - \phi_{ch} = (\phi_j - \phi_{ch})exp\left(-\frac{x-x_{Sj}}{\lambda_{ch}}\right)$. Then the average change in channel potential "$<U'>$" is given by:

$$<U'> = \frac{1}{(\phi_j-\phi_{ch})}\int_{\phi_{ch}}^{\phi_j} U' \, dU' = \frac{1}{(\phi_j-\phi_{ch})}\int_{\infty}^{x_{Sj}} U' \frac{\partial U'}{\partial x} dx = \frac{(\phi_j-\phi_{ch})}{2}, \qquad (A.7)$$

On the other hand, according to (A.5) the electric field is given by:

$$\frac{\partial U}{\partial x} = -\frac{(\phi_j-\phi_{ch})}{\lambda_{ch}} exp\left(-\frac{x-x_{Sj}}{\lambda_{ch}}\right), \qquad (A.8)$$

The average electric field "$<E_{ch}>$" inside the gated channel can then be calculated as follows:

$$<E_{ch}> = \frac{1}{(\phi_j-\phi_{ch})}\int_{\phi_{ch}}^{\phi_j} \frac{\partial U}{\partial x} dU = \frac{1}{(\phi_j-\phi_{ch})}\int_{\infty}^{x_{Sj}} \frac{\partial U}{\partial x}\frac{\partial U}{\partial x} dx = -\frac{(\phi_j-\phi_{ch})}{2\lambda_{ch}}, \qquad (A.9)$$

The distance associated with the band-bending inside the gated channel "$d_{ch}$" is assumed to be an average quantity defined by the ratio between $<U'_{ch}>$ and $<E_{ch}>$ (in magnitude). Therefore, according to (A.7) and (A.9):

$$d_{ch} = \left|\frac{<U'_{ch}>}{<E_{ch}>}\right| = \lambda_{ch}. \qquad (A.10)$$

Thus, the total band bending distance (see "$\Lambda$" in Fig. 1) is given by:

$$\Lambda = W_D + d_{ch} = W_D + \lambda_{ch}. \qquad (A.11)$$

Where $W_D$ is the depletion width inside the source region. Notice that this band bending distance as shown in (A.11) is approximately the same as the tunneling distance when the TFET device operates close to the threshold. On the other hand, the average electric field given by (A.5) can be calculated following a procedure analogous to that shown in (A.9):

$$<E> = \frac{1}{\phi_S-\phi_{ch}}\int_{\phi_S}^{\phi_{ch}} \frac{\partial U}{\partial x} dU = \frac{1}{\phi_S-\phi_{ch}}\int_{-\infty}^{+\infty} \frac{\partial U}{\partial x}\frac{\partial U}{\partial x} dx = -\frac{\phi_S-\phi_{ch}}{6\lambda_{eff}}, \qquad (A.12)$$

The denominator in the right hand side of (A.12) represents the average distance over which the total potential (i.e. $\phi_S - \phi_{ch}$) drops from source to channel when (A.5) is used to describe the potential profile. Hence, by combining (A.11) and (A.12) it is concluded that $6\lambda_{eff} = W_D + \lambda_{ch}$ and therefore the potential decay rate is given by:

$$\lambda_{eff} = \frac{W_D + \lambda_{ch}}{6}. \tag{A.13}$$

The value of $W_D$ is given by the classical expression:

$$W_D = \sqrt{\frac{2\epsilon}{qN_D}\alpha(\phi_S - \phi_{ch})}. \tag{A.14}$$

Where $\epsilon$ is the permittivity of the semiconductor, $N_D$ is the doping in the source and $\alpha$ is a number used to satisfy the condition $\alpha(\phi_S - \phi_{ch}) \approx (\phi_S - \phi_j)$ and thus $\alpha \approx (\phi_S - \phi_j)/(\phi_S - \phi_{ch})$. Although strictly speaking $\alpha$ depends on the conditions of the problem it is found to be close to a constant value which is $\alpha \approx 0.64$ and small deviations from this number do not change significantly the results since it has a squared root dependence. The expression shown in (A.13) is valid in the on-state of the device and is ultimately justified by the fact that it matches very well the results obtained from NGEF (as shown in the discussion). In the off-state where BTBT takes place directly between source and drain, the values of $\lambda_{eff}$ and $W_D$ are no longer relevant. Therefore, (A.13) is adequate both in the on- and off-state of the TFET.

### A.3 : Calculation of Maximum Electric Field

To find maximum magnitude of electric field ($\frac{\partial U}{\partial x}$), the second derivative of (A.5) is evaluated:

$$\frac{\partial^2 U(x)}{\partial x^2} = (\phi_S - \phi_{ch}) \frac{\exp\left(\frac{x+x_{Sj}}{\lambda_{eff}}\right)\left(\exp\left(\frac{x}{\lambda_{eff}}\right) - \exp\left(\frac{x_{Sj}}{\lambda_{eff}}\right)\right)}{\lambda_{eff}^2 \left(\exp\left(\frac{x_{Sj}}{\lambda_{eff}}\right) + \exp\left(\frac{x}{\lambda_{eff}}\right)\right)^3}, \tag{A.15}$$

At $x = x_{Sj}$ (A.15) is zero and thus:

$$|E_{Max}| = \left|\frac{\partial U}{\partial x}\right|_{x_{Sj}} = \frac{(\phi_S - \phi_{ch})}{4\lambda_{eff}}. \tag{A.16}$$

### A.4: Calculation Of The "Effective Channel Length" (*L*<sub>eff</sub>)

Equation (A.1) implies that the analytic potential profile is symmetrical with respect to the junction ($x = x_{Sj}$). However, the actual potential profile is not symmetric when $W_D \neq \lambda_{ch}$. In this case (where $W_D \neq \lambda_{ch}$), the symmetry point of potential shifts toward the source whenever $W_D$ is larger than $\lambda_{ch}$ and toward the gate otherwise. Notice that the effective channel length ($L_{eff}$) does not affect the on-state performance because most of the tunneling takes place around the junction ($x = x_{Sj}$) from source to channel (or channel to drain). However, in the off-state of the device most of the tunneling takes place directly from source to drain. Therefore, in the off-state the value of $L_{eff}$ is important and thus the actual position of the junctions (which ultimately determine the value of $L_{eff}$) must be properly defined. To account for this effect in equation (A.5), the symmetry point of the junction is shifted according to the difference of $W_{DS}$ and $\lambda_{ch}$:

$$\delta x_{Sj} \approx \frac{W_{DS}-\lambda_{ch}}{2}, \tag{A.17}$$

$$\delta x_{Dj} \approx \frac{W_{DD}-\lambda_{ch}}{2}, \tag{A.18}$$

Where $\delta x_{Sj}$ and $\delta x_{Dj}$ are the symmetry corrections at the source and drain junctions respectively, and $W_{DS}$ and $W_{DD}$ are the depletion widths corresponding to source and drain respectively. In the case of piecewise potential, it is empirically found that $\delta x_{Sj} = \delta x_{Dj} = \lambda_{ch}$. Notice that the shift is zero when the $W_{DS}$ and $\lambda_{ch}$ are equal. Thus the value of $L_{eff}$ can be taken as:

$$L_{eff} \approx L_{ch} + \delta x_{Sj} + \delta x_{Dj}. \tag{A.19}$$

### A.5: Analytic Description of Elliptic Complex Bands

In previous research work it has been demonstrated that the imaginary waevector denoted as "$k_{IM}(E)$" are described by an elliptic curve which can be approximated using a highly accurate analytic formula (within 1.4% error compared to rigorous tight-binding simulations)[39]. Such formula requires as input the values of the bandgap of the semiconductor $E_g$, and the corresponding hole and electron effective masses ($m_h$ and $m_e$ respectively). The analytic expression for $k_{IM}(E)$ is repeated here for the sake of clarity:

$$k_{IM}(E) = \frac{1}{\hbar}\sqrt{2m_h E \left(1 - \frac{E}{2E_q}\right)}, \qquad \text{for } 0 < E < E_q \qquad (A.20a)$$

$$k_{IM}(E) = \frac{1}{\hbar}\sqrt{2m_e (E_g - E)\left(1 - \frac{E_g - E}{2(E_g - E_q)}\right)}. \qquad \text{for } E_q < E < E_g \qquad (A.20b)$$

Where

$$E_q = E_g \frac{m_e}{m_h + m_e}. \qquad (A.20c)$$

In equation (A.20c) the value of $E_q$ is known as the "branching point" and it represents the point in energy inside the bandgap where the carrier changes from hole to electron behavior (or vice versa).

## A.6: General Applicability Of The Analytic Model For The Potential Profile

In order to show the general applicability of equation (6), Fig. A1 shows the comparison between the potential profiles obtained through self-consistent simulations and that obtained through (6) for a different channel material. The simulated device is a gate-all-around silicon nanowire TFET with aggressively scaled-down dimensions: (SiO$_2$ thickness) $t_{ox}$=1nm, $d$=4nm (diameter of the nanowire), $L_{ch}$ = 20nm. According to the results shown in Figs. 3 and A1 equation (6) seems to describe accurately the change in the potential along the channel for different $\phi_{ch}$, different doping concentrations, and for different materials when compared with the results obtained from rigorous NEGF simulations.

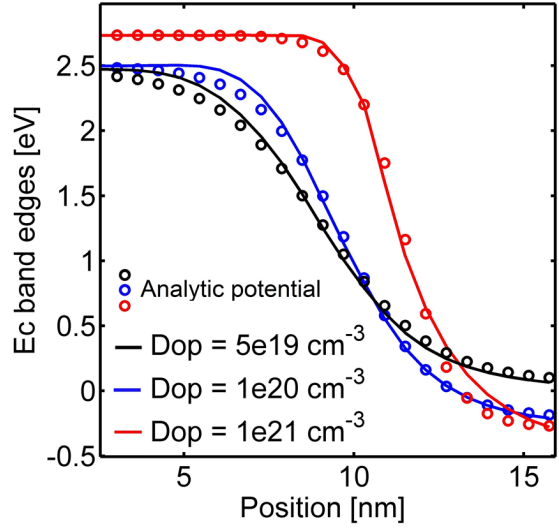

**Figure A1.** Conduction band potential profiles as a function of position along the channel. The simulated device is a Si nanowire gate-all around TFET. The source/channel interface is located at $x_{Sj}$=10nm. Solid lines are results obtained from NEGF and symbols are values calculated through equation (6).

**A.7: Spectral Current Obtained From The Compact Model**

Figure A2**a** shows the transfer characteristic obtained through equation (11) of a hypothetical ballistic device (see parameters in Fig. A2). Figures A2**b**-A2**f** show the spectral currents calculated through (10) corresponding band diagrams for some bias points indicated in Fig. A2**a**. The color bar represents the value of transmission probability for a particular value of energy, i.e., $T_{WKB}(E_\alpha)$.

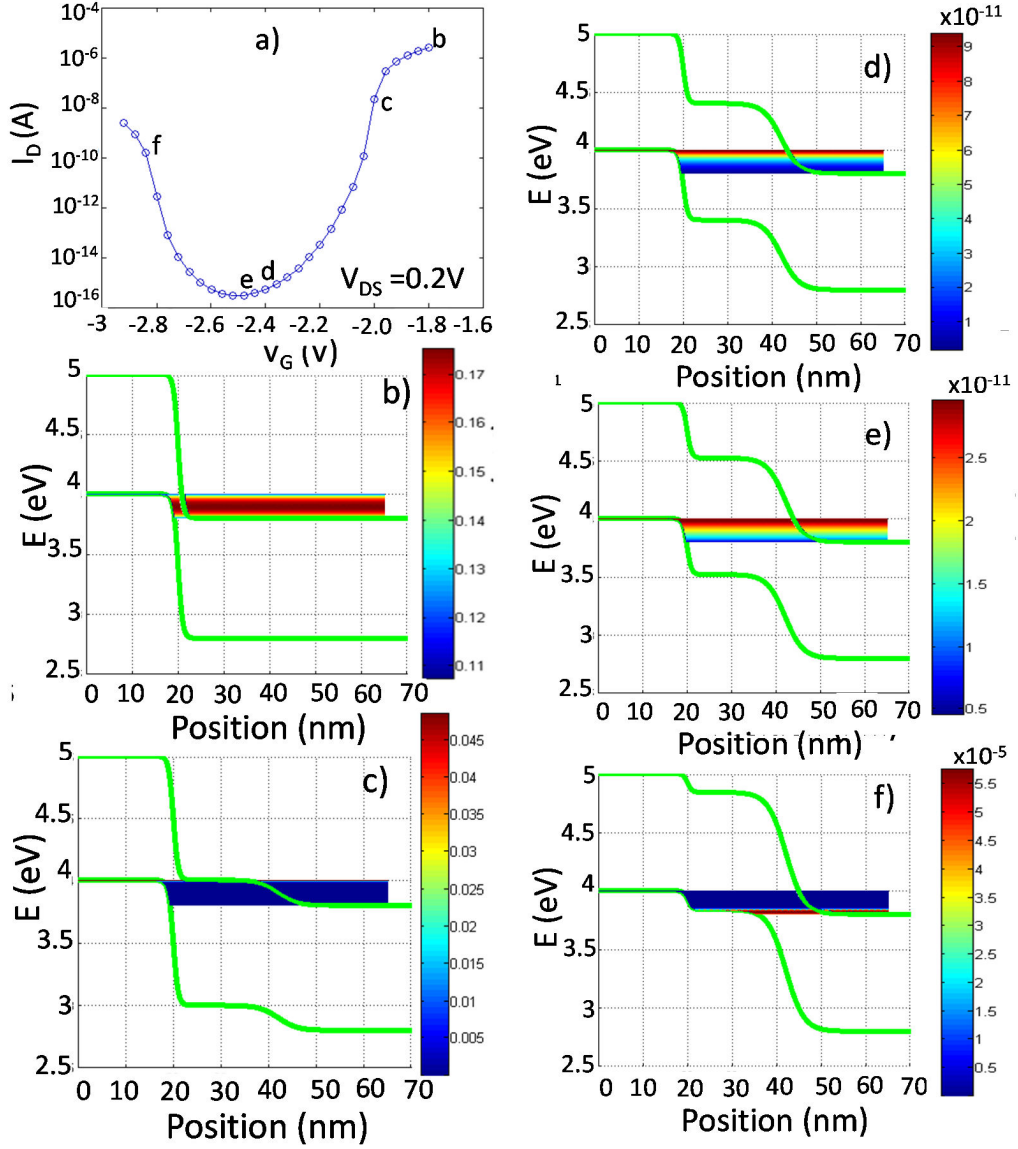

**Figure A2. a)** $I_D$-$V_G$ of a hypothetical ballistic TFET obtained through equation (11). All points between "**c**" and "**f**" correspond to direct tunneling from source to drain and they mimic the curvature of $k_{IM}(E)$. **b-f)** Band diagrams for the bias points indicated in A2**a** calculated with equation (10). The color bar represents the value of $T_{WKB}(E_\alpha)$ for a particular energy in the band diagram. For simplicity it is assumed that $(f_{E,S} - f_{E,D}) \cdot S = 1$. Simulation parameters: $m_h$ =0.038$m_0$, $m_e$ =0.04$m_0$, $E_g = 1eV$, $L_{ch} = 30nm$, $\lambda_{eff} = 2nm$, at the source and $\lambda_{eff} = 5nm$ at the drain.

Notice that the BTBT current values between points **b** and **c** is due to carriers tunneling from source to channel. All points between **"c"** and **"f"** correspond to direct tunneling from source to drain. This causes a dramatic deterioration of the subthreshold slope (*S*) deep in the off-state (especially close to points **d** and **e**). This is highly undesirable since a very steep *S* (lower than 60mV/decade) is one of the most attractive features of TFETs. To understand the underlying physics of the curvature of $I_D$-$V_G$ in the off-state consider the following: According to (11) and assuming low drain bias ($V_D$) the off-state current is approximated as $I_{off} \approx \frac{2q}{h} e^{-2k_{IM}(E_\alpha)L_{ch}} qV_D$. Where $E_\alpha$ is the energy at which carriers tunnel from source to drain. Notice that $E_\alpha \propto qV_G$ and therefore $I_{off}$ varies with $V_G$ mimicking the curvature of $k_{IM}(E)$. The minus sign in the exponential indicates that the higher $k_{IM}(E)$ the lower $I_{off}$ (and vice versa) but $I_{off}$ still resembles $k_{IM}(E)$. In this way it becomes clear that the upturn in the transfer characteristics between **e** and **f** is not related to turning on the channel/drain tunneling junction. Instead, the upturning is due to the decrease of $k_{IM}(E)$ corresponding to carriers tunneling from source to drain which reflects in the (increasing) magnitude of the BTBT current. Only in points beyond **f** the BTBT current is dominated by carriers tunneling from channel to drain. The insight that $I_{off}$ mimics the curvature of $k_{IM}(E)$ is especially relevant in a voltage-scaling trend where $V_D$ becomes small. In this case improving the off-state performance (reducing both *S* and $I_{off}$) requires to use semiconductors with large effective mass: since $I_{off}$ resembles $k_{IM}(E)$ and $k_{IM}(E) \propto \sqrt{m^*_{e,h}}$ then using large $m^*_{e,h}$ would translate into a $I_{off}$ that decreases faster with decreasing $V_G$ (sharper $I_D$-$V_G$ curvature). However, this would come at the expense of reducing the on-state current and thus improving both the off- and on-state performance requires further study in ultra-scaled TFETs operating at low bias.